\documentclass[twocolumn]{aastex631}

\usepackage{amsmath} % 
\usepackage{bm}

\usepackage{comment}
\usepackage{amsmath, amsthm, amssymb, amsfonts}
\usepackage{hyperref}
\usepackage{upgreek}
\usepackage{booktabs}
\usepackage{graphicx, nicefrac}
\usepackage{xcolor}

\defcitealias{Peng2010}{\scshape P10}
\defcitealias{Peng2012}{\scshape P12}

\begin{document}

%TC:ignore
\title{On the signature of black holes on the quenched stellar mass function}

\correspondingauthor{Antonio J. Porras-Valverde}
\email{antonio.porras@yale.edu}

\author[0000-0002-1996-0445]{Antonio J. Porras-Valverde}
\affiliation{Department of Astronomy, Yale University, P.O. Box 208101, New Haven, CT 06520, USA}

\author[0000-0002-1975-4449]{John C. Forbes}
\affiliation{School of Physical and Chemical Sciences--Te Kura Mat\=u, University of Canterbury, Christchurch 8140, New Zealand}

%%%%
%\author{C. D. Biemesderfer\altaffilmark{4,5}}
%\affil{National Optical Astronomy Observatories, Tucson, AZ 85719}
%\email{aastex-help@aas.org}

%\and

%\author{R. J. Hanisch\altaffilmark{5}}
%\affil{Space Telescope Science Institute, Baltimore, MD 21218}
%%%%


%%%%
%\altaffiltext{1}{Visiting Astronomer, Cerro Tololo Inter-American Observatory.
%CTIO is operated by AURA, Inc.\ under contract to the National Science
%\altaffiltext{2}{Society of Fellows, Harvard University.}
%\altaffiltext{3}{present address: Center for Astrophysics,
%    60 Garden Street, Cambridge, MA 02138}
%\altaffiltext{4}{Visiting Programmer, Space Telescope Science Institute}
%\altaffiltext{5}{Patron, Alonso's Bar and Grill}
%%%%


\begin{abstract}

As star-forming galaxies approach or exceed a stellar mass around $10^{11} M_\odot$, they are increasingly likely to be quenched in a process generically called mass quenching. Central galaxies, which are quenched via mass rather than environmental quenching, therefore accumulate in a peak around this characteristic mass. While a number of processes may influence the shape of the quenched central stellar mass function (QCSMF), we find that its low-mass slope is strongly affected by the scatter in the mass of black holes at a given stellar mass, with higher scatters in the black hole population yielding shallower slopes. Higher scatters in the black hole mass spread out the stellar mass range over which quenching occurs, leading to shallower slopes. This trend holds across a variety of semi-analytic models and cosmological hydrodynamic simulations. A comparison with observations provides indirect evidence for a large scatter in black hole mass $\sigma(\log_{10}(M_\mathrm{BH})|M_*) \gtrsim 0.5$ dex, and a joint constraint on AGN feedback physics and the co-evolution of galaxies and black holes.

\end{abstract}

%TC:endignore
\section{Introduction}\label{introduction}
%%%%%%%%%%%%%%%%%%%%%%

The connection between supermassive black holes and galaxies has become highly relevant in recent years. Observations demonstrate correlations between the mass of central massive black holes and properties of their host galaxies such as stellar mass, bulge mass, and velocity dispersion \citep{Magorrian1998, Ferrarese2000, Tremaine2002, Kormendy2013, Ding2020}. Generally, galaxies contain at least one massive black hole at their center, potentially driving significant outflows that disturb the gas availability depriving or enhancing galaxy 
and/or black hole growth. As \citet{SilkRees1998} suggested, the energy released during the gas accretion onto these black holes may be enough to disperse galactic gas and initiate outflows. This feedback mechanism from active galactic nuclei (AGN) is supported by substantial evidence from detection of X-ray cavities, radio jets in galaxy clusters, and Broad Absorption Lines in Quasi-Stellar Object (QSO) spectra \citep[e.g.][]{McNamara2007, Chon2012, Arav2013, Guo2015, Arav2018}.

Feedback mechanisms are believed to address several persistent challenges in cosmological simulations of galaxy formation, including over-cooling, low angular momentum, the existence of massive blue galaxies, and extra-galactic enrichment. Semi-analytical models of galaxy formation (SAMs) which combine N-body simulations of dark matter halos with analytical prescriptions of baryonic physics, have provided a partial solution to some of these issues by incorporating AGN feedback that prevents gas in the halo from accreting onto the galaxy, thereby allowing for the successful match of the bright-end of the galaxy luminosity function \citep{Bower2006, Croton2006, Somerville+2008, Somerville2015, Croton2016}.

\citet{Peng2010} and \citet{Peng2012} (hereafter \citetalias{Peng2010} and \citetalias{Peng2012} respectively) introduced a framework for understanding the stellar mass function of the galaxy population by splitting galaxies into quenched and star-forming populations, and requiring continuity, namely that star-forming galaxies become quenched galaxies. We refer to quenching as the processes that halt star formation in star-forming galaxies, leading to the emergence of passive galaxies with very low or zero star formation rates. This cessation of star formation, whether caused by internal or external factors, leads to the formation of the so-called ``red sequence" of passive galaxies. Quenching differs from the overall decline in the specific star formation rate of star-forming galaxies observed since $z\sim2$.

\citetalias{Peng2010} examined the implications of the observed distinct separation of the red fraction of galaxies in the Sloan Digital Sky Survey (SDSS) by mass and environment and the observed constancy of the characteristic $\mathcal{M}^*$ of the mass function of star-forming galaxies since $z\sim2$. The red fraction of galaxies indicates the proportion of galaxies that have been quenched at a given mass and/or in a specific environment. \citetalias{Peng2010} argued that this separability suggests two main quenching channels: one dependent on environment (environment quenching) and another linked to the mass of the galaxy (mass quenching). Mass quenching is independent of environment, and environment quenching is independent of stellar mass. \citetalias{Peng2012} further demonstrated that environment quenching is confined to satellite galaxies.

The \citetalias{Peng2010} framework shows clear relationships among the Schechter parameters of different components of the galaxy population, defined via \citet{schechter_1976},
\begin{equation}
    \Phi(M_*) d\log M_* = \Phi^* \left(\frac{M_*}{\mathcal{M}^*}\right)^{\alpha_s+1} \exp\left(-\frac{M_*}{\mathcal{M}^*}\right) d\log M_*,
\end{equation}
where $\Phi(M_*)$ is the number density of galaxies per unit log interval in stellar mass, $\alpha_s$ is the low-mass slope, $\mathcal{M}^*$ is the characteristic mass, and $\Phi^*$ is the normalization. Note that for a Schechter function when $M_*\ll \mathcal{M}^*$, $\alpha\equiv d\log\Phi/d\log M_*=\alpha_s+1$. If the mass quenching process that produces the exponential cutoff in the mass function of star-forming galaxies extends to masses below $\mathcal{M}^*$, then the faint-end slope of mass-quenched passive galaxies is predicted to differ from that of star-forming galaxies by $\Delta \alpha_s \sim 1+\beta$, where $\beta\approx -0.1$ is the powerlaw slope of the specific star formation rate vs. stellar mass \citepalias{Peng2010}. A major success of the \citetalias{Peng2010} framework is that this relationship holds in the data. Meanwhile the faint-end slope of environment-quenched galaxies should align with that of the star-forming population, as it does in fact in the data. This leads to a distinctive double Schechter function for the overall population of passive galaxies. Similarly, a double Schechter function is also expected for the mass function of the entire galaxy population \citep{Peng2010, Baldry2012}.

Given the apparent correspondence between mass quenching and the physics of AGN feedback, the details of which are still highly uncertain, we explore how the extension of mass quenching to masses below $\mathcal{M}^*$ works in a variety of simulations and SAMs. \citetalias{Peng2012} showed that observationally the data are well-described by a gradual mass quenching process in which the population of passive central galaxies is built up over a wide range of stellar masses, producing a Schechter low-mass slope $\alpha_{s,\mathrm{red}}\approx -0.33$. While simulations and SAMs are typically tuned to reproduce the overall stellar mass function (along with other quantities) reasonably well, the low-mass slope of the mass function of quenched central galaxies is typically not directly tuned. This slope is nonetheless highly sensitive to the details of mass quenching, including the physics of AGN feedback, and the black hole population responsible.

This paper is structured as follows: Section \ref{Methodology} provides an overview of the semi-analytic model {\sc Dark Sage} and the modifications applied in our study. Section \ref{SMF_plot} presents the SMF for both all and central galaxies. Section \ref{model_comparisons} analyzes the SMF of central galaxies, distinguishing between star-forming and quenched galaxies, with quenched galaxies defined as those with $\log_{10}(\mathrm{sSFR}\ [\text{yr}^{-1}]) < -11$. Section \ref{SMF_Q_alpha} explains how we obtain $\alpha$ from the SMF of quenched galaxies. In Section \ref{alpha_SMFQ_constrain}, we explore the relationship between $\alpha$ and $\sigma_{\rm BH}$. Finally, Section \ref{discussion_conclusion} summarizes the conclusions of the study. The Appendix \ref{SMF_Q_cen_allmodels} includes detailed SMFs and $\alpha$ fits for all models.

\section{Summary of models}\label{Methodology}

We use the semi-analytic model {\sc Dark Sage} \citep{Stevens2016}, an extension of the Semi-Analytic Galaxy Evolution ({\sc SAGE}) model which uses physically motivated prescriptions to model gas cooling, star formation, and feedback mechanisms \citep{Croton2016}. What sets {\sc Dark Sage} apart from many other SAMs is its incorporation of a self-consistent 1-dimensional disk structure, divided into 30 annuli with fixed angular momentum. We adopt black hole formation model modifications to approximate results from the {\sc IllustrisTNG300-1} simulation \citep{PorrasValverde2024}. In the SMBH seeding model, we assign a $1.1 \times 10^{6}\, \mathrm{M}{\odot}$ black hole to every halo that reaches $7  \times  10^{10}\, \mathrm{M}{\odot}$, in contrast to {\sc Dark Sage}'s fiducial approach of using low-mass SMBH seeds. We also experiment with a ``fixed conditional distribution model," in which all black hole masses follow a predetermined distribution at a fixed halo mass, centered on the median black hole mass–halo mass relation at $z=0$ from {\sc IllustrisTNG}, with an imposed lognormal scatter at fixed halo mass $\sigma_\mathrm{BH}$. By imposing a fixed relationship between halo mass and black hole mass, we can separate out the effect of changes in the underlying black hole population from changes in the feedback prescription. However, in doing so we decouple the typical relationships between black hole mass and AGN feedback. Three different AGN feedback models are used in the SMBH seeding and fixed conditional distribution models:

\begin{itemize}
\item Instantaneous AGN Feedback: {\sc Dark Sage} calculates its fiducial black hole accretion rates and energy outputs given the current mass of the black hole.
\item Black Holes Turn Off Cooling: Deactivates radiative cooling for any galaxy with a black hole mass greater than $10^8 M_\odot$, with no additional AGN feedback mechanisms.
\item Black Holes Remove ISM: Cold Interstellar Medium (ISM) gas from galaxies with black hole masses exceeding $10^{8} M_\odot$ is removed.
\end{itemize}

In addition to our SAM models, we compare to predictions from the {\sc UniverseMachine} semi-empirical model, which uses cosmological N-body simulations to statistically match observed galaxy properties across cosmic time \citep{Behroozi2019Universemachine:010}, and the {\sc SHARK} SAM of galaxy formation \citep{LagosSHARK2024}. We include the {\sc IllustrisTNG} \citep{Pillepich2018, Springel2018, Naiman2018, Nelson2018, Marinacci2018, Nelson2019} and {\sc EAGLE} \citep{2016A&C....15...72M} state-of-the-art cosmological hydrodynamic simulations used to study astrophysical processes related to star formation, cold gas availability, black hole formation, stellar evolution, stellar and AGN feedback. 

\section{Galaxy SMF} \label{SMF_plot}

\begin{figure}[hbt!]
\centering
\includegraphics[width=\columnwidth, clip]{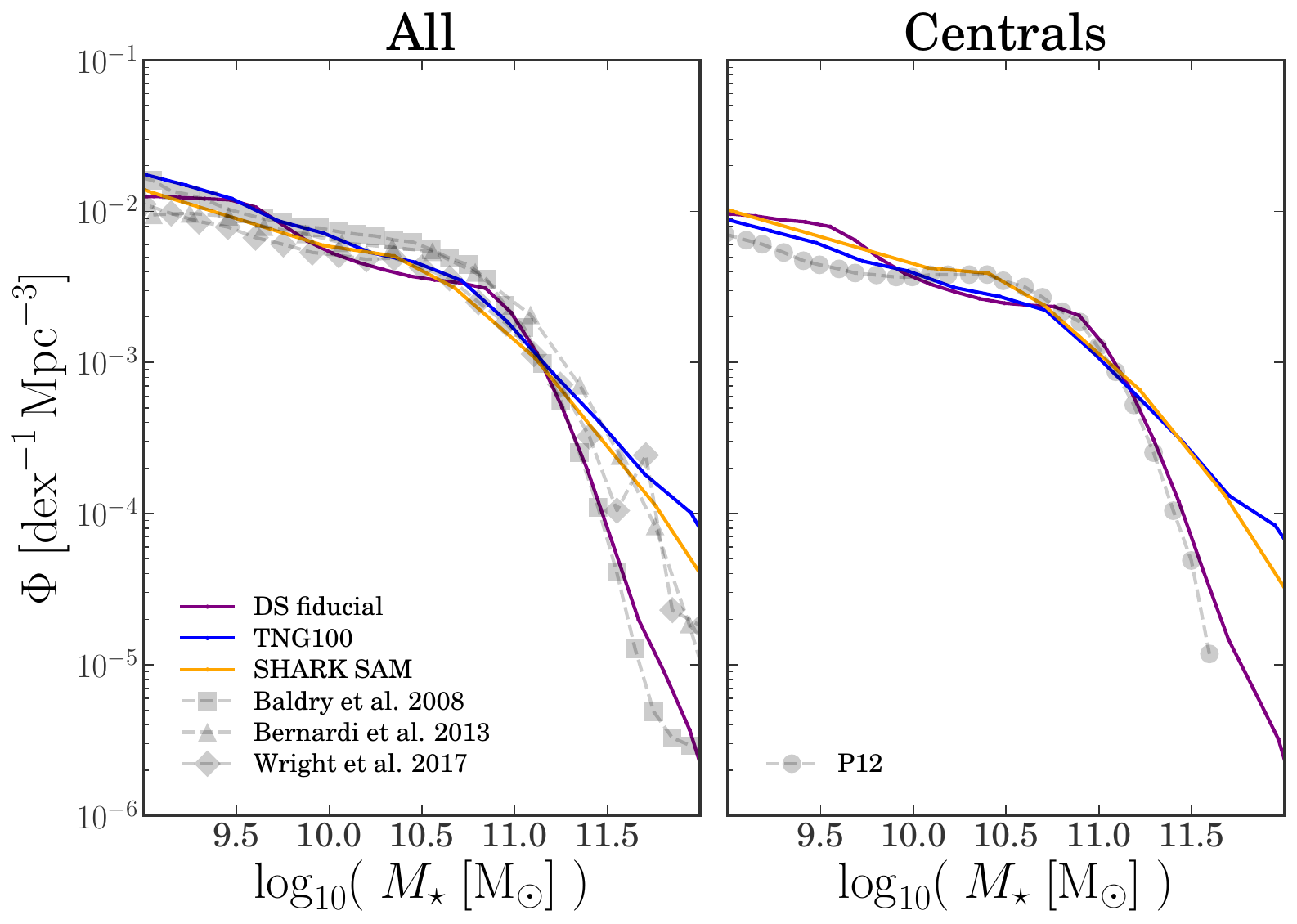}
\caption{Galaxy SMF for all (left panel) and central (right panel) galaxies at $z=0$. Both panels show {\sc Dark Sage} (solid purple line), {\sc TNG100} (solid blue line), and {\sc SHARK} (solid orange line) models. We include observations from \citet{Baldry2008} (grey squares dashed line), \citet{Bernardi2013} (grey triangles dashed line), and \citep{Wright2017} (black dashed dotted line) (grey diamonds dashed line), and \citetalias{Peng2012} (grey circles dashed line) (right panel). We find agreement within all models with increasing orders of magnitude uncertainty at the high-mass end.}\label{fig:SMF_z0}

\end{figure}

Before delving\footnote{No artificial neural networks were used in writing this work.} into the quenched central stellar mass function, we compare the stellar mass function (SMF) of all galaxies at $z=0$ to various models. Simulations are generally tuned to provide plausible matches to the SMF at $z=0$. While inferring the stellar mass of galaxies does come with systematic uncertainty, it is among the most accessible observable properties of a galaxy \citep{pacifici_2023}. We also compare the mass functions of central galaxies between the observations and the models. In the models, the definition of a central galaxy is unambiguous, but the observations require a model-dependent group catalog to classify galaxies as centrals or satellites. 

Figure \ref{fig:SMF_z0} shows the SMF for all and central galaxies at $z=0$. Overall, {\sc Dark Sage}, {\sc TNG100}, and {\sc SHARK} models align with observations from \citet{Baldry2008}, \citet{Bernardi2013}, and \citep{Wright2017} within $0.6\, \mathrm{dex}$. However, discrepancies increase to $1.5\, \mathrm{dex}$ for $M_* > 10^{11.5} M_\odot$ (left panel). The agreement between models and observations from \citetalias{Peng2012} remain consistent when considering only central galaxies (right panel). Note that {\sc Dark Sage}\footnote{We caution the reader in interpreting low-mass galaxy results. Although {\sc Dark Sage} is tuned to observations, galaxies with $M_* \lesssim 10^{10} M_\odot$ have less than 200 particles in their present dark matter halo.} is calibrated to match the SMF from \citet{Baldry2008}, {\sc TNG100} with the SMF from \citet{Bernardi2013}, and {\sc SHARK} with the SMF from \citet{Wright2017}. Therefore, the nature of the agreement is not a coincidence. The disagreement at the high-mass end is apparent in observations due to the uncertainty in the mass budget from intra-cluster light not consistently accounted in the total stellar mass budget \citep{Bernardi2013}. At fixed high-mass, {\sc TNG100} and {\sc SHARK} show a higher number density of galaxies when compared to {\sc Dark Sage}. 

\section{Star-forming and Quenched galaxy SMF} \label{model_comparisons}

\begin{figure*}[t]
\centering
\includegraphics[width=2.0\columnwidth, clip]{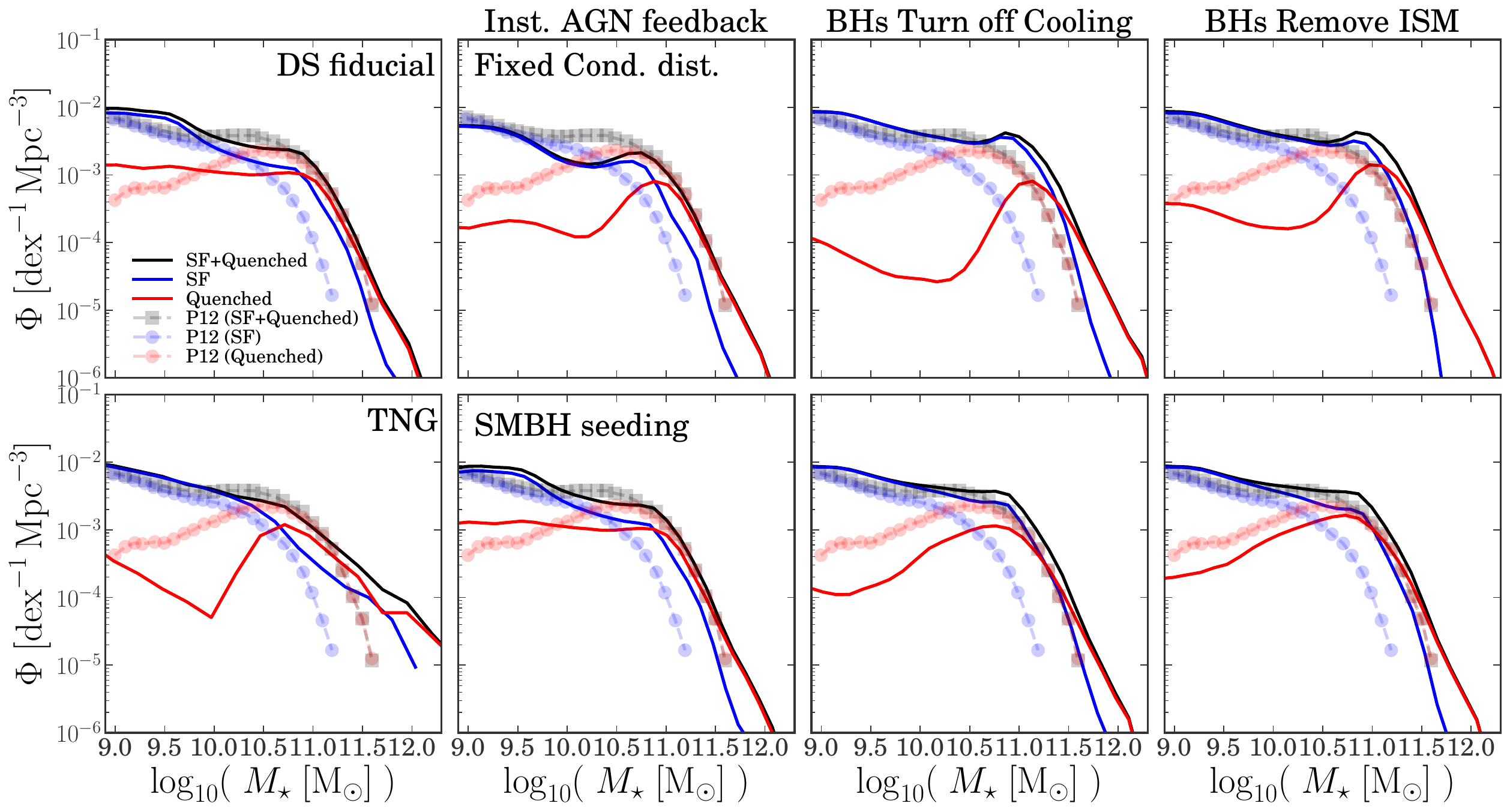}
\caption{SMF of central star-forming ($-11 < \log_{10}(\mathrm{sSFR}\ [\text{yr}^{-1}])$, solid blue line) and quenched galaxies ($-11 > \log_{10}(\mathrm{sSFR}\ [\text{yr}^{-1}])$, solid red line) at $z=0$ for {\sc Dark Sage} (DS) fiducial (top left), {\sc TNG100} (bottom left), fixed conditional distribution models (top three panels), and SMBH seeding models (bottom three panels) using the \textit{instantaneous AGN feedback} (first column), \textit{black holes turn off cooling} (second column), and \textit{black holes remove ISM} (last column) quenching models. Note that the fixed conditional distribution model we show here is for a fixed $\sigma_{\rm BH} = 0.2$. We show \citetalias{Peng2012}'s centrals galaxies (solid black line) broken down into star-forming (solid blue line) and quenched (solid red line) galaxies in each of the panels. While the SMF of star-forming galaxies steadily declines with mass (in most models) just like the SMF of centrals overall, central quenched galaxies show a wide range of behaviors below the characteristic mass of the star-forming galaxies.}\label{fig:SMF_SFQuenched_models_z0}

\end{figure*}

We now split galaxies into star-forming and quenched populations at a specific star formation rate of $10^{-11}\ \mathrm{yr}^{-1}$, low enough that quenched galaxies can no longer adjust their mass significantly in a Hubble time via star formation alone. Figure \ref{fig:SMF_SFQuenched_models_z0} shows the SMF for central galaxies split into all, star-forming, and quenched galaxies. First, \citetalias{Peng2012}'s passive central galaxy population is described by a single Schechter function with a {\it positive} faint-end slope, leading to a well-defined peak around $\mathcal{M}^*$. The characteristic stellar mass $\mathcal{M}^*$ is $0.09 \pm 0.03\, \mathrm{dex}$ larger than that of star-forming galaxies, likely due to minor post-quenching mergers of satellite galaxies into the centrals, which boosts their masses by about 25\% on average \citepalias{Peng2012}.

The star-forming populations across models are reasonably consistent with each other, although none are as featureless as the data from \citetalias{Peng2012}. The offset in stellar mass at a fixed number density reaches $0.25-0.6\, \mathrm{dex}$ at the high-mass end depending on the model. In all cases, at $M_* \gtrsim 10^{11.5} M_\odot$, quenched galaxies make up the vast majority of galaxies (i.e. the red and black lines follow each other), except in {\sc TNG100} where we see a slight difference in number density. 

Despite the general consistency of the star-forming galaxies, the models do have features around $\mathcal{M}^*$ that are not present in the data. In particular, each version of {\sc Dark Sage} has a knee or even a bump in the SMF of star-forming galaxies. This over-abundance of star-forming galaxies appears to originate in a star-forming main sequence whose normalization is somewhat low in these versions of {\sc Dark Sage}. Galaxies that have not yet quenched have such low specific star formation rates that they are unable to grow in stellar mass (and hence move rightwards in the SMF). This freeze-out of star-forming galaxies can happen when the specific star formation rate of the main sequence decreases below $\sim 1/t_H$, which evidently has not happened in the real Universe, but could happen in the future as galaxies continue to become more gas-poor at a given stellar mass.

While the features of the star-forming SMF are interesting, the clearer signature of quenching is present in the quenched galaxies. All of the models have a clear feature, usually a peak in the QCSMF around $10^{11} M_\odot$, which is understood in \citet{Peng2010} as the result of a mass-quenching process at this mass scale. All models need to have this feature in order to reproduce the total SMF. Despite the agreement in the location of the peak and its behavior above $\mathcal{M}^*$, there is a huge variety of behaviors below $\mathcal{M}^*$ in the QCSMF: flat or negative slopes, positive slopes, and positive slopes that become negative again at lower masses. Since we have already removed satellite galaxies from consideration, these behaviors reflect differences in the mass-quenching process below $\mathcal{M}^*$. The QCSMF well below $\mathcal{M}^*$ in the models may also be influenced by the effects of finite resolution (e.g. a galaxy that ``should'' be near the main sequence, but momentarily has no star-forming cells), or by the tail of galaxies $\sim 3\sigma$ below the main sequence. While such galaxies are rare, there are so many more star-forming galaxies at these masses than quenched central galaxies that they may account for the increase in the QCSMF at very low stellar masses. Few if any of the models' QCSMF can be represented with a single Schechter function like the \citetalias{Peng2012} data\footnote{The \citetalias{Peng2012} data below their completeness limit actually exceeds their best-fit Schechter function, so there is some indication the observations also have an upturn at very low stellar masses.}, so we proceed by measuring the slope of the QCSMF in the regime just below $\mathcal{M}^*$ which can plausibly be represented by a single powerlaw value.

\section{Obtaining $\alpha$}\label{SMF_Q_alpha}

\begin{figure*}[t]
\centering
\includegraphics[width=2.0\columnwidth, clip]{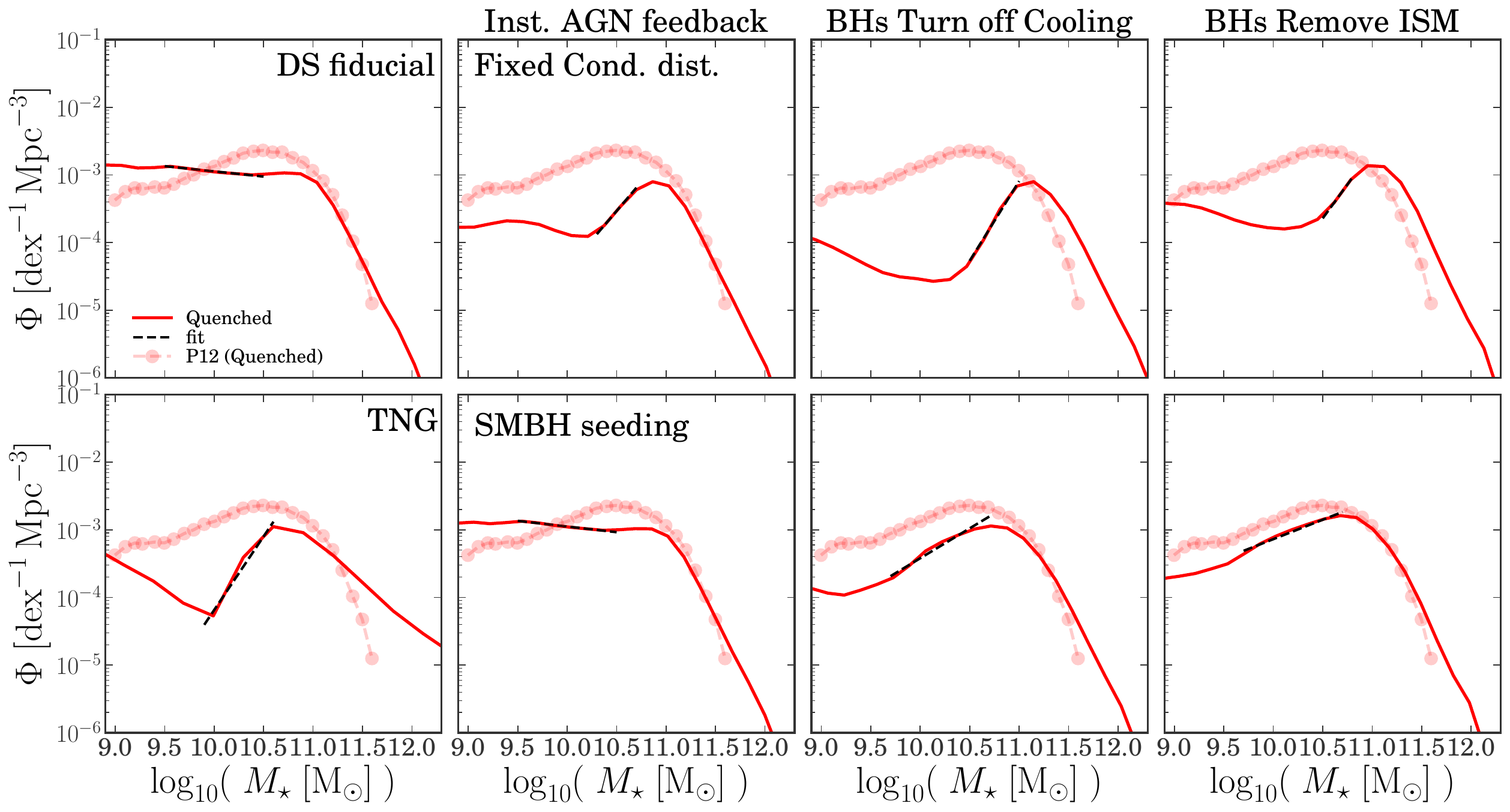}
\caption{Similar to figure \ref{fig:SMF_SFQuenched_models_z0}, but here we only show quenched galaxies for {\sc TNG100}, {\sc Dark Sage} fiducial and its modified models. The black dashed line shows the low-end fit to obtain the slope $\alpha$. We highlight differences in $\alpha$ with varying BH growth and quenching models.}\label{fig:SMF_Q_models_z0}

\end{figure*}

Figure \ref{fig:SMF_Q_models_z0} illustrates how we determine the faint-end slope $\alpha$ of the QCSMF. The stellar mass range over which we measure $\alpha$ is chosen by hand for each model as the largest values of $M_*$ below the peak in the QCSMF where the slope is plausibly represented by a single powerlaw value. We observe that $\alpha$ varies significantly across different numerical models. In the {\sc Dark Sage} fiducial model, the transition between star-forming and quenched galaxies is less-abrupt than the {\sc TNG100} model \citep[see fig. 5 of][]{PorrasValverde2024}. This results in a flatter faint-end slope for the QCSMF in the {\sc Dark Sage} fiducial model compared to {\sc TNG100}. The slope presented by \citetalias{Peng2012} lies between these two models. We investigate how various modifications to {\sc Dark Sage} impact $\alpha$. In the fixed conditional distribution models, the $\sigma_\mathrm{BH} = 0.2$ model shows a steeper slope for all quenching scenarios compared to the SMBH seeding case, which exhibits some variation in $\alpha$ depending on the quenching models. These variations in $\alpha$ depending on the AGN feedback prescription appear to be secondary to variations in $\alpha$ due to differences in the black hole population (see next section).

\section{Constraining AGN feedback using $\alpha$}\label{alpha_SMFQ_constrain}

\begin{figure*}[t]
\centering
\includegraphics[width=1.5\columnwidth, clip]{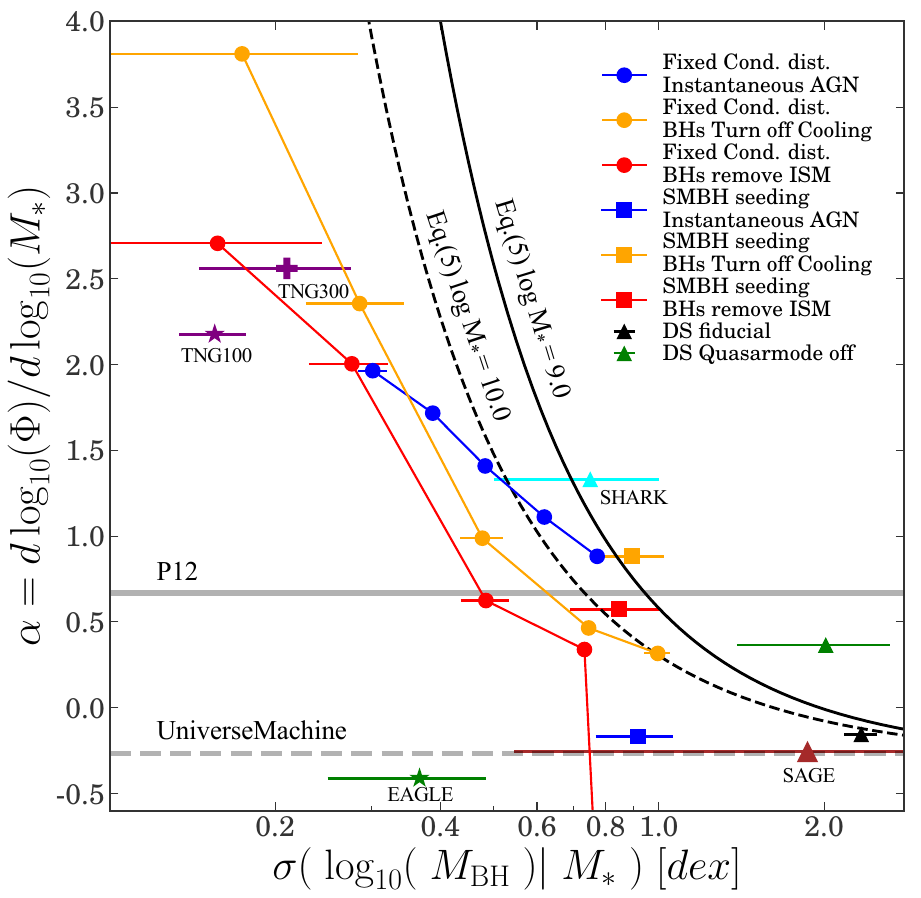}
\caption{Faint-end slope $\alpha$ of the SMF of central quenched galaxies as a function of the black hole mass scatter at fixed stellar mass. Horizontal error bars show the span of scatters within the stellar mass range used to estimate $\alpha$. We show the following models: {\sc Dark Sage} fiducial (black triangle), {\sc Dark Sage} with \textit{quasar-mode} feedback turned off (green triangle), the fixed conditional distribution (circles), and the SMBH seeding (squares) models. For both {\sc Dark Sage} modifications, we include the \textit{instantaneous AGN feedback} (blue), \textit{black holes turn off cooling} (orange), and \textit{black holes remove ISM} (red) quenching models. We also present data from the {\sc UniverseMachine} empirical model (grey dashed line), cosmological hydrodynamic simulations {\sc TNG100} (purple star) and {\sc EAGLE} (green star), and SAMs {\sc SAGE} (brown triangle) and {\sc SHARK} (cyan triangle). The observational constraint from \citetalias{Peng2012} (solid grey line) is shown for comparison. We overplot our prediction from a toy model where black holes fully quench their galaxies when they reach a particular mass (equation \ref{eq:dlogfQdlogMstar}) for several stellar masses: $M_* \sim 10^{9} M_\odot$ (solid black line) and $M_* \sim 10^{10} M_\odot$ (dashed black line). We find a correlation where increasing $\sigma_\mathrm{BH}$ yields lower $\alpha$ values.}\label{fig:alpha_sigmaBH}

\end{figure*}

Figure \ref{fig:alpha_sigmaBH} shows the slope $\alpha$ of the faint-end of the SMF of central quenched galaxies as a function of $\sigma_\mathrm{BH}$, the scatter in black hole mass at fixed stellar mass. Generally, galaxies with lower $\sigma_\mathrm{BH}$ values yield higher $\alpha$. This holds between simulations, and within a given simulation where we modify the underlying physics. The values of $\alpha$ derived from the simulations are compared to \citetalias{Peng2012}'s value of $\alpha$ from their single Schechter function fit (note that we add 1 to account for comparison with the logarithmic slopes presented here). The {\sc UniverseMachine}'s value of $\alpha$, an output of their empirical model, differs from \citetalias{Peng2012}'s more direct measurement of $\alpha$, an issue discussed further in the next section.

\section{Discussion \& Conclusions}\label{discussion_conclusion}

Using a modified version of the SAM {\sc Dark Sage} where we can easily adjust the scatter in the black hole mass at a fixed galaxy mass, we have found a clear trend that higher scatter in the black hole mass leads to lower values of the low-mass slope of the stellar mass function for quenched central galaxies. This trend also holds for other SAMs ({\sc SAGE} and {\sc SHARK}) and hydrodynamic simulations ({\sc TNG100} and {\sc EAGLE}). This relationship is intuitive -- larger scatter in the black hole mass implies a wider range of stellar masses over which quenching occurs, and therefore a shallower slope in the population of galaxies quenched by this process.

To reproduce the observed low-mass slope in the mass function of central quenched galaxies, our trends suggest a reasonably large scatter in the black hole mass at fixed stellar mass is necessary, with $\sigma_\mathrm{BH} \ga 0.5$ dex at the very least. Shallow slopes in line with observations may also be possible for physical models of AGN feedback that are more gradual than any implemented in our versions of {\sc Dark Sage}, or in {\sc TNG} for that matter. In most cases, these models of AGN feedback work by invoking a dramatic increase in the energy available to heat galactic halos for black holes above a critical mass around $10^8 M_\odot$. It is conceivable that the same population of black holes could produce sufficiently gradual mass quenching if the feedback turned on more gradually with black hole mass. However, in our models that come closest to this case, namely the ``Instantaneous AGN'' models, the $\sigma_\mathrm{BH}$ necessary to produce the observed $\alpha$ is even higher, approaching 1 dex. This is because in these models the halo mass continues to play a large role in quenching.

This presents a problem for physical models of galaxy formation, both cosmological simulations and SAMs, since the scatter in black hole mass is not an input, but rather an emergent property of the simulation. While it would be straightforward to modify the AGN feedback recipe, it is much less obvious how to induce the model to increase its scatter in black hole mass. The growth of the black holes apparently needs to be less well-coupled to the growth of stellar mass in its host galaxy.

The intrinsic scatter in the $M_{\rm BH}-M_*$ and $M_{\rm BH}-M_{\rm bulge}$ relations is a key measure of black hole–galaxy coevolution. \citet{Kormendy2013} and \citet{Bennert2021} find a smaller scatter of 0.28–0.39 dex when focusing only on classical bulges and pseudobulges. However, when all morphological types and both active and inactive galaxies are included, the scatter increases to about 0.5 dex for the $M_{\rm BH}-M_{\rm bulge}$ relation and slightly more for the $M_{\rm BH}-M_*$ relation \citep{Reines&Volonteri2015, Bentz2018}. The $M_{\rm BH}-M_*$ relation for local massive black holes from \citet{Greene2020} shows lower normalization and a scatter of $\sim$0.81 dex, which is based on a combined fit of local early- and late-type galaxies with mostly dynamical black hole mass measurements. \citet{2023ApJ...954..173L} show an intrinsic scatter of $0.59^{+0.23}_{-0.21}$ dex for the $M_{\rm BH}-M_{\rm bulge}$ relation and $0.47^{+0.24}_{-0.17}$ dex for the $M_{\rm BH}-M_*$ relation, which aligns with the observed local BH scaling relations. Here, they correct for the \citet{Lauer2007} bias, a statistical effect caused by intrinsic scatter in scaling relations within flux-limited samples. \citet{2015MNRAS.452..575S} found a tight $M_{\rm BH}-M_*$ relation at $z=0$ for massive black holes, with increasing scatter for those below  
$10^{8} M_\odot$, similar to SDSS observations from \citep{2023ApJ...954..173L}. This variation in scatter between high- and low-mass black holes may indicate different evolutionary processes or feedback mechanisms shaping the black hole scaling relations.

Some measures of Reverberation Mapping black hole mass estimates may be underestimated if the intrinsic scatter in the virial factor for individual objects is not accounted for (e.g., $\sim$0.3–0.4 dex; \citet{2024ApJS..272...26S}). High-resolution Hubble Space Telescope images may struggle to resolve bulges and other galaxy structures precisely, introducing further uncertainties in bulge identification and mass estimation. Consequently, if $M_{\rm BH}$ and $M_*$ uncertainties are underestimated, the observed intrinsic scatter in scaling relations might be overestimated, as the ``true" scatter is obscured by measurement errors. Taken together the observational constraints on the value of $\sigma(\log_{10} M_\mathrm{BH}|M_*)$ are consistent with our inference that its value is likely to be $\gtrsim 0.5$ dex based on the observed value of $\alpha$. 

We now address the relationship between our view here, that $\sigma_\mathrm{BH}$ controls $\alpha$, and the \citetalias{Peng2010} result that, at low stellar masses,
\begin{equation} \label{eq:pengalpha}
    \alpha = \alpha_\mathrm{blue} + 1 + \beta
\end{equation}
where $\alpha_\mathrm{blue}$ is the low-mass slope of the star-forming mass function, and $1+\beta$ is the powerlaw slope of the star-forming main sequence. The \citetalias{Peng2010} formula is derived from requiring that star-forming galaxies follow a single Schechter function with a non-evolving characteristic mass, which requires that a galaxy's probability of quenching during a given period of time, $\eta$, is proportional to its star formation rate. This argument only applies near the characteristic mass $\mathcal{M}^* \approx 10^{11} M_\odot$, but \citetalias{Peng2010} point out that this simple quenching law, if extended to lower masses, predicts Equation \ref{eq:pengalpha}, which in fact is numerically satisfied for the observed galaxy population at $z\approx 0$.

In contrast, many of the models we consider here have a quenching probability that is not directly related to their star formation rate, but rather to their black hole mass. Our {\it black holes remove ISM} model is the most extreme of these, guaranteeing that galaxies containing black holes above $10^8 M_\odot$ have identically zero star formation. In such a model, we can write down the quenched fraction given a stellar mass as
\begin{equation} \label{eq:ncdf}
    f_Q =  \mathrm{NCDF}\left( \frac{\log M_\mathrm{BH,med}(M_*)- \log M_{\mathrm{BH,crit}} }{\sigma_\mathrm{BH}} \right)
\end{equation}
Here NCDF is the cumulative distribution function of a standard Gaussian, $M_\mathrm{BH,crit} \sim 10^8 M_\odot$ is the critical black hole mass above which a galaxy is quenched, and $M_\mathrm{BH,med}(M_*)$ is the median black hole mass at a given stellar mass. We can further assume that $M_\mathrm{BH,med} = A (M_*/10^{11} M_\odot)^\gamma$ with $\gamma \sim 1$ and $A\sim 10^{8} M_\odot$. 

Now let us consider how our $f_Q$ interacts with \citetalias{Peng2010}'s apparently successful ansatz that $\eta = \mathrm{SFR}/\mathcal{M}^*$ at all masses. At masses well below $\mathcal{M}^*$, we can find $f_Q$ as predicted by the \citetalias{Peng2010} formalism just by noting that the density of star-forming galaxies in this regime $\Phi_\mathrm{SF}$ is far larger than that of quenched galaxies $\Phi_Q$, in which case
\begin{equation}
    f_Q = \frac{\Phi_Q}{\Phi_Q+\Phi_\mathrm{SF}} \approx \frac{\Phi_Q}{\Phi_\mathrm{SF}} \propto M_*^{1+\beta}.
\end{equation}
Meanwhile we can Taylor expand the value of $\log f_Q$ based on sudden black hole quenching (Equation \ref{eq:ncdf}) to find
\begin{equation}
\label{eq:dlogfQdlogMstar}
    \frac{d\log f_Q}{d\log M_*} =  \frac{2 \gamma}{\ln(10)}  \left(\frac{1}{\sigma_\mathrm{BH} \sqrt{2\pi}} - \gamma \frac{\log M_*- y}{\pi  \sigma_\mathrm{BH}^2}\right) 
\end{equation}
where $y=11\gamma-\log A + \log M_\mathrm{BH,crit}$. We can compare this prediction directly to our measured values of $\alpha$ by noting that in general $d \log f_Q/d\log M_* = \alpha-\alpha_\mathrm{blue}$. 
If we require that this slope approaches the \citetalias{Peng2010} limit of $1+\beta$ at a particular $M_*$, we obtain a quadratic equation for $\sigma_\mathrm{BH}$, 
\begin{equation}
    \sigma_\mathrm{BH} = \frac{1+\sqrt{1-4\ln(10)(1+\beta)( \log M_*-y)}}{\sqrt{2\pi} \ln(10)(1+\beta) / \gamma}.
\end{equation}
For $\gamma=1$, $A=M_\mathrm{BH,crit}=10^8 M_\odot$, $\beta=-0.1$, and $M_* = 10^{10} M_\odot$, we find $\sigma_\mathrm{BH}\approx 0.78$ dex, roughly what we find for the simulations that match the \citetalias{Peng2012} slope.

In this way a quenching law based purely on instantaneous black hole mass can mimic the single Schechter function of quenched central galaxies observed in \citetalias{Peng2012}. We can see this explicitly in Figure \ref{fig:SMF_SFQuenched_models_z0} and the Appendix, where the \textit{black holes remove ISM} models, especially those with large $\sigma_\mathrm{BH}$ produce QCSMFs with a shape quite similar to the observations\footnote{The location of the characteristic mass is also sensitive to $\sigma_\mathrm{BH}$, but this can be modified by a suitable choice of $M_{\rm BH,crit}.$}. It is difficult to distinguish between quenching mechanisms observationally, where there is some evidence for the importance of black holes \citep[e.g.][]{Terrazas_2017}, but the halo mass or the central density of the galaxy may also play a role \citep{Fang_2013, Woo_2015}. In simulations, however, black hole activity appears to be the only viable means to halt star formation. The success of \citetalias{Peng2010}'s mass quenching model wherein the quenching probability is proportional to the star formation rate, may be a coincidence: the star formation rate does not matter \textit{per se} -- more massive galaxies have both higher star formation rates and a higher probability of hosting a black hole above the critical mass. For this probability to be high enough at low enough stellar masses, $\sigma_\mathrm{BH}$ must be reasonably large, $\gtrsim 0.5$ dex.

To summarize,
\begin{itemize}
    \item Increasing $\sigma_\mathrm{BH}$ is correlated with shallower slopes in QCSMF (lower $\alpha$ values). Larger $\sigma_\mathrm{BH}$ spreads the ``mass quenching'' process over a larger range of stellar masses, which appears to be necessary to explain the shallow slope.
    \item The value of $\sigma_\mathrm{BH}$ necessary to produce a realistic slope, $\gtrsim 0.5$ dex, is consistent with efforts to measure $\sigma_\mathrm{BH}$ directly.
\end{itemize}

 %TC:ignore

\software{IPython \citep{Perez2007IPythonFor}, 
    Scipy \citep{2020SciPy-NMeth}, 
    matplotlib \citep{Hunter2007MatplotlibEnvironment}, 
    Astropy \citep{Robitaille2013Astropy:Astronomy}, 
    NumPy \citep{VanDerWalt2011TheComputation}
         }

\vspace{10pt}
%\begin{acknowledgments}
AJPV delightedly acknowledges support from the Heising-Simons foundation. We thank Priya Natarajan, Pieter van Dokkum, Meg Urry, Frank van den Bosch, Shy Genel, Lars Hernquist, Jenny Greene, and Rachel Somerville for helpful discussions and comments. JCF is grateful for support from the New Zealand Government, administered by the Royal Society Te Ap\={a}rangi. We used computational facilities from the Vanderbilt Advanced Computing Center for Research and Education (ACCRE) and the Yale Center for Research Computing (YCRC). Literature reviews for this work were made using the NASA’s Astrophysics Data System. 
%\end{acknowledgments}

\clearpage

\bibliographystyle{aasjournal}
\bibliography{bibfile.bib}

\begin{thebibliography}{}
\expandafter\ifx\csname natexlab\endcsname\relax\def\natexlab#1{#1}\fi
\providecommand{\url}[1]{\href{#1}{#1}}
\providecommand{\dodoi}[1]{doi:~\href{http://doi.org/#1}{\nolinkurl{#1}}}
\providecommand{\doeprint}[1]{\href{http://ascl.net/#1}{\nolinkurl{http://ascl.net/#1}}}
\providecommand{\doarXiv}[1]{\href{https://arxiv.org/abs/#1}{\nolinkurl{https://arxiv.org/abs/#1}}}

\bibitem[{{Arav} {et~al.}(2013){Arav}, {Borguet}, {Chamberlain}, {Edmonds}, \& {Danforth}}]{Arav2013}
{Arav}, N., {Borguet}, B., {Chamberlain}, C., {Edmonds}, D., \& {Danforth}, C. 2013, \mnras, 436, 3286, \dodoi{10.1093/mnras/stt1812}

\bibitem[{{Arav} {et~al.}(2018){Arav}, {Liu}, {Xu}, {Stidham}, {Benn}, \& {Chamberlain}}]{Arav2018}
{Arav}, N., {Liu}, G., {Xu}, X., {et~al.} 2018, \apj, 857, 60, \dodoi{10.3847/1538-4357/aab494}

\bibitem[{Baldry {et~al.}(2008)Baldry, Glazebrook, \& Driver}]{Baldry2008}
Baldry, I.~K., Glazebrook, K., \& Driver, S.~P. 2008, MNRAS, 959, 945, \dodoi{10.1111/j.1365-2966.2008.13348.x}

\bibitem[{{Baldry} {et~al.}(2012){Baldry}, {Driver}, {Loveday}, {Taylor}, {Kelvin}, {Liske}, {Norberg}, {Robotham}, {Brough}, {Hopkins}, {Bamford}, {Peacock}, {Bland-Hawthorn}, {Conselice}, {Croom}, {Jones}, {Parkinson}, {Popescu}, {Prescott}, {Sharp}, \& {Tuffs}}]{Baldry2012}
{Baldry}, I.~K., {Driver}, S.~P., {Loveday}, J., {et~al.} 2012, \mnras, 421, 621, \dodoi{10.1111/j.1365-2966.2012.20340.x}

\bibitem[{Behroozi {et~al.}(2019)Behroozi, Wechsler, Hearin, \& Conroy}]{Behroozi2019Universemachine:010}
Behroozi, P., Wechsler, R.~H., Hearin, A.~P., \& Conroy, C. 2019, Monthly Notices of the Royal Astronomical Society, 488, 3143, \dodoi{10.1093/mnras/stz1182}

\bibitem[{{Bennert} {et~al.}(2021){Bennert}, {Treu}, {Ding}, {Stomberg}, {Birrer}, {Snyder}, {Malkan}, {Stephens}, \& {Auger}}]{Bennert2021}
{Bennert}, V.~N., {Treu}, T., {Ding}, X., {et~al.} 2021, \apj, 921, 36, \dodoi{10.3847/1538-4357/ac151a}

\bibitem[{{Bentz} \& {Manne-Nicholas}(2018)}]{Bentz2018}
{Bentz}, M.~C., \& {Manne-Nicholas}, E. 2018, \apj, 864, 146, \dodoi{10.3847/1538-4357/aad808}

\bibitem[{{Bernardi} {et~al.}(2013){Bernardi}, {Meert}, {Sheth}, {Vikram}, {Huertas-Company}, {Mei}, \& {Shankar}}]{Bernardi2013}
{Bernardi}, M., {Meert}, A., {Sheth}, R.~K., {et~al.} 2013, Monthly Notices of the Royal Astronomical Society, 436, 697, \dodoi{10.1093/mnras/stt1607}

\bibitem[{{Bower} {et~al.}(2006){Bower}, {Benson}, {Malbon}, {Helly}, {Frenk}, {Baugh}, {Cole}, \& {Lacey}}]{Bower2006}
{Bower}, R.~G., {Benson}, A.~J., {Malbon}, R., {et~al.} 2006, \mnras, 370, 645, \dodoi{10.1111/j.1365-2966.2006.10519.x}

\bibitem[{{Chon} {et~al.}(2012){Chon}, {B{\"o}hringer}, {Krause}, \& {Tr{\"u}mper}}]{Chon2012}
{Chon}, G., {B{\"o}hringer}, H., {Krause}, M., \& {Tr{\"u}mper}, J. 2012, \aap, 545, L3, \dodoi{10.1051/0004-6361/201219538}

\bibitem[{Croton {et~al.}(2006)Croton, Springel, White, De~Lucia, Frenk, Gao, Jenkins, Kauffmann, Navarro, \& Yoshida}]{Croton2006}
Croton, D.~J., Springel, V., White, S.~D., {et~al.} 2006, Monthly Notices of the Royal Astronomical Society, 365, 11, \dodoi{10.1111/j.1365-2966.2005.09675.x}

\bibitem[{{Croton} {et~al.}(2016){Croton}, {Stevens}, {Tonini}, {Garel}, {Bernyk}, {Bibiano}, {Hodkinson}, {Mutch}, {Poole}, \& {Shattow}}]{Croton2016}
{Croton}, D.~J., {Stevens}, A. R.~H., {Tonini}, C., {et~al.} 2016, The Astrophysical Journal Supplement Series, 222, 22, \dodoi{10.3847/0067-0049/222/2/22}

\bibitem[{{Ding} {et~al.}(2020){Ding}, {Silverman}, {Treu}, {Schulze}, {Schramm}, {Birrer}, {Park}, {Jahnke}, {Bennert}, {Kartaltepe}, {Koekemoer}, {Malkan}, \& {Sanders}}]{Ding2020}
{Ding}, X., {Silverman}, J., {Treu}, T., {et~al.} 2020, The Astrophysical Journal, 888, 37, \dodoi{10.3847/1538-4357/ab5b90}

\bibitem[{{Fang} {et~al.}(2013){Fang}, {Faber}, {Koo}, \& {Dekel}}]{Fang_2013}
{Fang}, J.~J., {Faber}, S.~M., {Koo}, D.~C., \& {Dekel}, A. 2013, \apj, 776, 63, \dodoi{10.1088/0004-637X/776/1/63}

\bibitem[{{Ferrarese} \& {Merritt}(2000)}]{Ferrarese2000}
{Ferrarese}, L., \& {Merritt}, D. 2000, \apjl, 539, L9, \dodoi{10.1086/312838}

\bibitem[{{Greene} {et~al.}(2020){Greene}, {Strader}, \& {Ho}}]{Greene2020}
{Greene}, J.~E., {Strader}, J., \& {Ho}, L.~C. 2020, \araa, 58, 257, \dodoi{10.1146/annurev-astro-032620-021835}

\bibitem[{{Guo}(2015)}]{Guo2015}
{Guo}, F. 2015, \apj, 803, 48, \dodoi{10.1088/0004-637X/803/1/48}

\bibitem[{Hunter(2007)}]{Hunter2007MatplotlibEnvironment}
Hunter, J.~D. 2007, Computing In Science and Engineering, 9

\bibitem[{{Kormendy} \& {Ho}(2013)}]{Kormendy2013}
{Kormendy}, J., \& {Ho}, L.~C. 2013, \araa, 51, 511, \dodoi{10.1146/annurev-astro-082708-101811}

\bibitem[{{Lagos} {et~al.}(2024){Lagos}, {Bravo}, {Tobar}, {Obreschkow}, {Power}, {Robotham}, {Proctor}, {Hansen}, {Chandro-G{\'o}mez}, \& {Carrivick}}]{LagosSHARK2024}
{Lagos}, C. d.~P., {Bravo}, M., {Tobar}, R., {et~al.} 2024, \mnras, 531, 3551, \dodoi{10.1093/mnras/stae1024}

\bibitem[{{Lauer} {et~al.}(2007){Lauer}, {Tremaine}, {Richstone}, \& {Faber}}]{Lauer2007}
{Lauer}, T.~R., {Tremaine}, S., {Richstone}, D., \& {Faber}, S.~M. 2007, \apj, 670, 249, \dodoi{10.1086/522083}

\bibitem[{{Li} {et~al.}(2023){Li}, {Shen}, {Ho}, {Brandt}, {Grier}, {Hall}, {Homayouni}, {Koekemoer}, {Schneider}, \& {Trump}}]{2023ApJ...954..173L}
{Li}, J. I.~H., {Shen}, Y., {Ho}, L.~C., {et~al.} 2023, \apj, 954, 173, \dodoi{10.3847/1538-4357/acddda}

\bibitem[{{Magorrian} {et~al.}(1998){Magorrian}, {Tremaine}, {Richstone}, {Bender}, {Bower}, {Dressler}, {Faber}, {Gebhardt}, {Green}, {Grillmair}, {Kormendy}, \& {Lauer}}]{Magorrian1998}
{Magorrian}, J., {Tremaine}, S., {Richstone}, D., {et~al.} 1998, \aj, 115, 2285, \dodoi{10.1086/300353}

\bibitem[{{Marinacci} {et~al.}(2018){Marinacci}, {Vogelsberger}, {Pakmor}, {Torrey}, {Springel}, {Hernquist}, {Nelson}, {Weinberger}, {Pillepich}, {Naiman}, \& {Genel}}]{Marinacci2018}
{Marinacci}, F., {Vogelsberger}, M., {Pakmor}, R., {et~al.} 2018, \mnras, 480, 5113, \dodoi{10.1093/mnras/sty2206}

\bibitem[{{McAlpine} {et~al.}(2016){McAlpine}, {Helly}, {Schaller}, {Trayford}, {Qu}, {Furlong}, {Bower}, {Crain}, {Schaye}, {Theuns}, {Dalla Vecchia}, {Frenk}, {McCarthy}, {Jenkins}, {Rosas-Guevara}, {White}, {Baes}, {Camps}, \& {Lemson}}]{2016A&C....15...72M}
{McAlpine}, S., {Helly}, J.~C., {Schaller}, M., {et~al.} 2016, Astronomy and Computing, 15, 72, \dodoi{10.1016/j.ascom.2016.02.004}

\bibitem[{{McNamara} \& {Nulsen}(2007)}]{McNamara2007}
{McNamara}, B.~R., \& {Nulsen}, P.~E.~J. 2007, \araa, 45, 117, \dodoi{10.1146/annurev.astro.45.051806.110625}

\bibitem[{{Naiman} {et~al.}(2018){Naiman}, {Pillepich}, {Springel}, {Ramirez-Ruiz}, {Torrey}, {Vogelsberger}, {Pakmor}, {Nelson}, {Marinacci}, {Hernquist}, {Weinberger}, \& {Genel}}]{Naiman2018}
{Naiman}, J.~P., {Pillepich}, A., {Springel}, V., {et~al.} 2018, \mnras, 477, 1206, \dodoi{10.1093/mnras/sty618}

\bibitem[{{Nelson} {et~al.}(2018){Nelson}, {Pillepich}, {Springel}, {Weinberger}, {Hernquist}, {Pakmor}, {Genel}, {Torrey}, {Vogelsberger}, {Kauffmann}, {Marinacci}, \& {Naiman}}]{Nelson2018}
{Nelson}, D., {Pillepich}, A., {Springel}, V., {et~al.} 2018, \mnras, 475, 624, \dodoi{10.1093/mnras/stx3040}

\bibitem[{{Nelson} {et~al.}(2019){Nelson}, {Pillepich}, {Springel}, {Pakmor}, {Weinberger}, {Genel}, {Torrey}, {Vogelsberger}, {Marinacci}, \& {Hernquist}}]{Nelson2019}
---. 2019, \mnras, 490, 3234, \dodoi{10.1093/mnras/stz2306}

\bibitem[{{Pacifici} {et~al.}(2023){Pacifici}, {Iyer}, {Mobasher}, {da Cunha}, {Acquaviva}, {Burgarella}, {Calistro Rivera}, {Carnall}, {Chang}, {Chartab}, {Cooke}, {Fairhurst}, {Kartaltepe}, {Leja}, {Ma{\l}ek}, {Salmon}, {Torelli}, {Vidal-Garc{\'\i}a}, {Boquien}, {Brammer}, {Brown}, {Capak}, {Chevallard}, {Circosta}, {Croton}, {Davidzon}, {Dickinson}, {Duncan}, {Faber}, {Ferguson}, {Fontana}, {Guo}, {Haeussler}, {Hemmati}, {Jafariyazani}, {Kassin}, {Larson}, {Lee}, {Mantha}, {Marchi}, {Nayyeri}, {Newman}, {Pandya}, {Pforr}, {Reddy}, {Sanders}, {Shah}, {Shahidi}, {Stevans}, {Triani}, {Tyler}, {Vanderhoof}, {de la Vega}, {Wang}, \& {Weston}}]{pacifici_2023}
{Pacifici}, C., {Iyer}, K.~G., {Mobasher}, B., {et~al.} 2023, \apj, 944, 141, \dodoi{10.3847/1538-4357/acacff}

\bibitem[{{Peng} {et~al.}(2012){Peng}, {Lilly}, {Renzini}, \& {Carollo}}]{Peng2012}
{Peng}, Y.-j., {Lilly}, S.~J., {Renzini}, A., \& {Carollo}, M. 2012, \apj, 757, 4, \dodoi{10.1088/0004-637X/757/1/4}

\bibitem[{{Peng} {et~al.}(2010){Peng}, {Lilly}, {Kova{\v{c}}}, {Bolzonella}, {Pozzetti}, {Renzini}, {Zamorani}, {Ilbert}, {Knobel}, {Iovino}, {Maier}, {Cucciati}, {Tasca}, {Carollo}, {Silverman}, {Kampczyk}, {de Ravel}, {Sanders}, {Scoville}, {Contini}, {Mainieri}, {Scodeggio}, {Kneib}, {Le F{\`e}vre}, {Bardelli}, {Bongiorno}, {Caputi}, {Coppa}, {de la Torre}, {Franzetti}, {Garilli}, {Lamareille}, {Le Borgne}, {Le Brun}, {Mignoli}, {Perez Montero}, {Pello}, {Ricciardelli}, {Tanaka}, {Tresse}, {Vergani}, {Welikala}, {Zucca}, {Oesch}, {Abbas}, {Barnes}, {Bordoloi}, {Bottini}, {Cappi}, {Cassata}, {Cimatti}, {Fumana}, {Hasinger}, {Koekemoer}, {Leauthaud}, {Maccagni}, {Marinoni}, {McCracken}, {Memeo}, {Meneux}, {Nair}, {Porciani}, {Presotto}, \& {Scaramella}}]{Peng2010}
{Peng}, Y.-j., {Lilly}, S.~J., {Kova{\v{c}}}, K., {et~al.} 2010, \apj, 721, 193, \dodoi{10.1088/0004-637X/721/1/193}

\bibitem[{P{\'{e}}rez \& Granger(2007)}]{Perez2007IPythonFor}
P{\'{e}}rez, F., \& Granger, B.~E. 2007, IEEE Journals {\&} Magazines, 9, 21, \dodoi{10.1109/MCSE.2007.53}

\bibitem[{{Pillepich} {et~al.}(2018){Pillepich}, {Nelson}, {Hernquist}, {Springel}, {Pakmor}, {Torrey}, {Weinberger}, {Genel}, {Naiman}, {Marinacci}, \& {Vogelsberger}}]{Pillepich2018}
{Pillepich}, A., {Nelson}, D., {Hernquist}, L., {et~al.} 2018, \mnras, 475, 648, \dodoi{10.1093/mnras/stx3112}

\bibitem[{{Porras-Valverde} {et~al.}(2024){Porras-Valverde}, {Forbes}, {Somerville}, {Stevens}, {Holley-Bockelmann}, {Berlind}, \& {Genel}}]{PorrasValverde2024}
{Porras-Valverde}, A.~J., {Forbes}, J.~C., {Somerville}, R.~S., {et~al.} 2024, \apj, 976, 148, \dodoi{10.3847/1538-4357/ad7b0f}

\bibitem[{{Reines} \& {Volonteri}(2015)}]{Reines&Volonteri2015}
{Reines}, A.~E., \& {Volonteri}, M. 2015, \apj, 813, 82, \dodoi{10.1088/0004-637X/813/2/82}

\bibitem[{Robitaille {et~al.}(2013)Robitaille, Tollerud, Greenfield, Droettboom, Bray, Aldcroft, Davis, Ginsburg, Price-Whelan, Kerzendorf, Conley, Crighton, Barbary, Muna, Ferguson, Grollier, Parikh, Nair, G{\"{u}}nther, Deil, Woillez, Conseil, Kramer, Turner, Singer, Fox, Weaver, Zabalza, Edwards, Azalee~Bostroem, Burke, Casey, Crawford, Dencheva, Ely, Jenness, Labrie, Lim, Pierfederici, Pontzen, Ptak, Refsdal, Servillat, \& Streicher}]{Robitaille2013Astropy:Astronomy}
Robitaille, T.~P., Tollerud, E.~J., Greenfield, P., {et~al.} 2013, Astronomy and Astrophysics, 558, 33, \dodoi{10.1051/0004-6361/201322068}

\bibitem[{{Schechter}(1976)}]{schechter_1976}
{Schechter}, P. 1976, \apj, 203, 297, \dodoi{10.1086/154079}

\bibitem[{{Shen} {et~al.}(2024){Shen}, {Grier}, {Horne}, {Stone}, {Li}, {Yang}, {Homayouni}, {Trump}, {Anderson}, {Brandt}, {Hall}, {Ho}, {Jiang}, {Petitjean}, {Schneider}, {Tao}, {Donnan}, {AlSayyad}, {Bershady}, {Blanton}, {Bizyaev}, {Bundy}, {Chen}, {Davis}, {Dawson}, {Fan}, {Greene}, {Gr{\"o}ller}, {Guo}, {Ibarra-Medel}, {Jiang}, {Keenan}, {Kollmeier}, {Lejoly}, {Li}, {de la Macorra}, {Moe}, {Nie}, {Rossi}, {Smith}, {Tee}, {Weijmans}, {Xu}, {Yue}, {Zhou}, {Zhou}, \& {Zou}}]{2024ApJS..272...26S}
{Shen}, Y., {Grier}, C.~J., {Horne}, K., {et~al.} 2024, \apjs, 272, 26, \dodoi{10.3847/1538-4365/ad3936}

\bibitem[{{Sijacki} {et~al.}(2015){Sijacki}, {Vogelsberger}, {Genel}, {Springel}, {Torrey}, {Snyder}, {Nelson}, \& {Hernquist}}]{2015MNRAS.452..575S}
{Sijacki}, D., {Vogelsberger}, M., {Genel}, S., {et~al.} 2015, Monthly Notices of the Royal Astronomical Society, 452, 575, \dodoi{10.1093/mnras/stv1340}

\bibitem[{{Silk} \& {Rees}(1998)}]{SilkRees1998}
{Silk}, J., \& {Rees}, M.~J. 1998, \aap, 331, L1, \dodoi{10.48550/arXiv.astro-ph/9801013}

\bibitem[{{Somerville} {et~al.}(2008){Somerville}, {Hopkins}, {Cox}, {Robertson}, \& {Hernquist}}]{Somerville+2008}
{Somerville}, R.~S., {Hopkins}, P.~F., {Cox}, T.~J., {Robertson}, B.~E., \& {Hernquist}, L. 2008, Monthly Notices of the Royal Astronomical Society, 391, 481, \dodoi{10.1111/j.1365-2966.2008.13805.x}

\bibitem[{{Somerville} {et~al.}(2015){Somerville}, {Popping}, \& {Trager}}]{Somerville2015}
{Somerville}, R.~S., {Popping}, G., \& {Trager}, S.~C. 2015, Monthly Notices of the Royal Astronomical Society, 453, 4337, \dodoi{10.1093/mnras/stv1877}

\bibitem[{{Springel} {et~al.}(2018){Springel}, {Pakmor}, {Pillepich}, {Weinberger}, {Nelson}, {Hernquist}, {Vogelsberger}, {Genel}, {Torrey}, {Marinacci}, \& {Naiman}}]{Springel2018}
{Springel}, V., {Pakmor}, R., {Pillepich}, A., {et~al.} 2018, \mnras, 475, 676, \dodoi{10.1093/mnras/stx3304}

\bibitem[{Stevens {et~al.}(2016)Stevens, Croton, \& Mutch}]{Stevens2016}
Stevens, A. R.~H., Croton, D.~J., \& Mutch, S.~J. 2016, Monthly Notices of the Royal Astronomical Society, 461, 859, \dodoi{10.1093/mnras/stw1332}

\bibitem[{{Terrazas} {et~al.}(2017){Terrazas}, {Bell}, {Woo}, \& {Henriques}}]{Terrazas_2017}
{Terrazas}, B.~A., {Bell}, E.~F., {Woo}, J., \& {Henriques}, B. M.~B. 2017, \apj, 844, 170, \dodoi{10.3847/1538-4357/aa7d07}

\bibitem[{{Tremaine} {et~al.}(2002){Tremaine}, {Gebhardt}, {Bender}, {Bower}, {Dressler}, {Faber}, {Filippenko}, {Green}, {Grillmair}, {Ho}, {Kormendy}, {Lauer}, {Magorrian}, {Pinkney}, \& {Richstone}}]{Tremaine2002}
{Tremaine}, S., {Gebhardt}, K., {Bender}, R., {et~al.} 2002, \apj, 574, 740, \dodoi{10.1086/341002}

\bibitem[{Van Der~Walt {et~al.}(2011)Van Der~Walt, Colbert, \& Varoquaux}]{VanDerWalt2011TheComputation}
Van Der~Walt, S., Colbert, S.~C., \& Varoquaux, G. 2011, Computing in Science and Engineering, 13, 22, \dodoi{10.1109/MCSE.2011.37}

\bibitem[{Virtanen {et~al.}(2020)Virtanen, Gommers, Oliphant, Haberland, Reddy, Cournapeau, Burovski, Peterson, Weckesser, Bright, {van der Walt}, Brett, Wilson, Millman, Mayorov, Nelson, Jones, Kern, Larson, Carey, Polat, Feng, Moore, {VanderPlas}, Laxalde, Perktold, Cimrman, Henriksen, Quintero, Harris, Archibald, Ribeiro, Pedregosa, {van Mulbregt}, \& {SciPy 1.0 Contributors}}]{2020SciPy-NMeth}
Virtanen, P., Gommers, R., Oliphant, T.~E., {et~al.} 2020, Nature Methods, 17, 261, \dodoi{10.1038/s41592-019-0686-2}

\bibitem[{{Woo} {et~al.}(2015){Woo}, {Dekel}, {Faber}, \& {Koo}}]{Woo_2015}
{Woo}, J., {Dekel}, A., {Faber}, S.~M., \& {Koo}, D.~C. 2015, \mnras, 448, 237, \dodoi{10.1093/mnras/stu2755}

\bibitem[{{Wright} {et~al.}(2017){Wright}, {Robotham}, {Driver}, {Alpaslan}, {Andrews}, {Baldry}, {Bland-Hawthorn}, {Brough}, {Brown}, {Colless}, {da Cunha}, {Davies}, {Graham}, {Holwerda}, {Hopkins}, {Kafle}, {Kelvin}, {Loveday}, {Maddox}, {Meyer}, {Moffett}, {Norberg}, {Phillipps}, {Rowlands}, {Taylor}, {Wang}, \& {Wilkins}}]{Wright2017}
{Wright}, A.~H., {Robotham}, A.~S.~G., {Driver}, S.~P., {et~al.} 2017, \mnras, 470, 283, \dodoi{10.1093/mnras/stx1149}

\end{thebibliography}

\appendix

%% ------------------------------------------------------------------------------------ %%  
   
%% ------------------------------------------------------------------------------------ %%  
\section{A. SMF of central star-forming and quenched galaxies for SMBH seeding and all fixed conditional distribution models at $z=0$} 
	\label{SMF_Q_cen_allmodels}

% say something to introduce the appendix

\begin{figure*}[t]
\centering
\includegraphics[width=\columnwidth, clip]{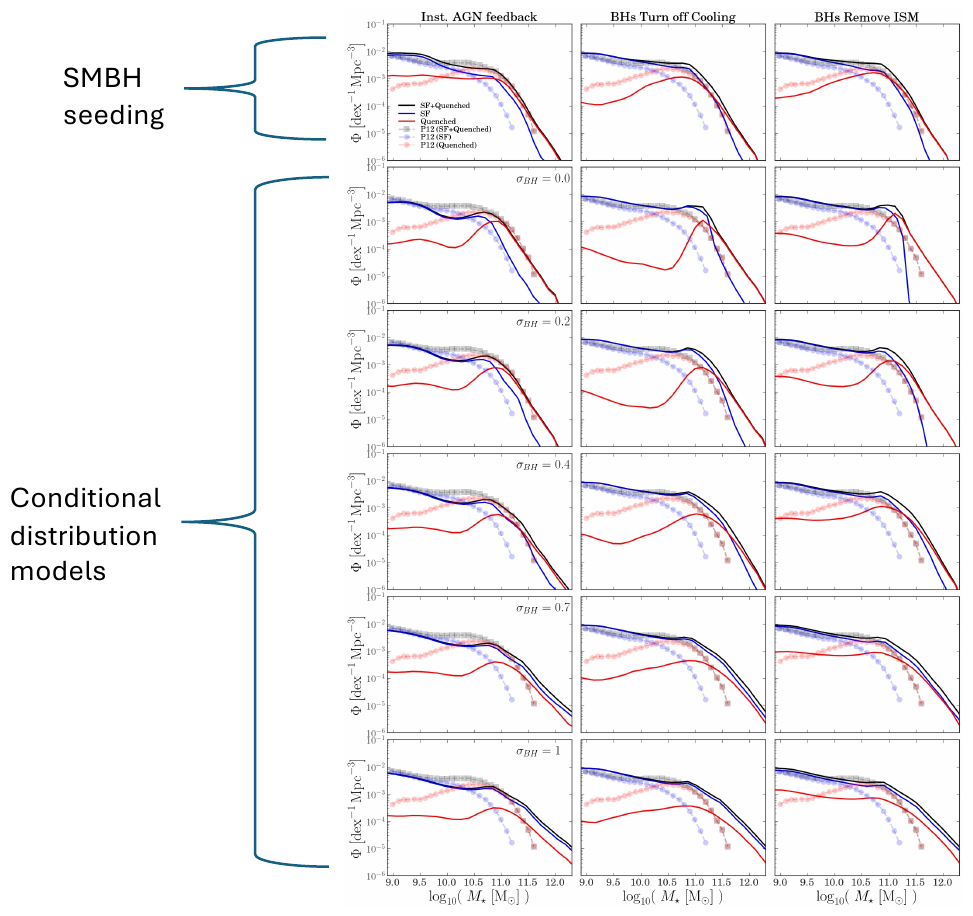}
\caption{SMF of central star-forming ($-11 < \log_{10}(\mathrm{sSFR}\ [\text{yr}^{-1}])$, solid blue line) and quenched galaxies ($-11 > \log_{10}(\mathrm{sSFR}\ [\text{yr}^{-1}])$, solid red line) at $z=0$ for both {\sc Dark Sage} modifications: the SMBH seeding models (top three panels) and the fixed conditional distribution models with varying $\sigma_{\rm BH}$ values from $\sigma_{\rm BH}=0$ (second row) to $\sigma_{\rm BH}=1$ (last row). Each column represents three different quenching models: the \textit{instantaneous AGN feedback} (first column), \textit{black holes turn off cooling} (second column), and \textit{black holes remove ISM} (last column) quenching models. We show \citetalias{Peng2012}'s centrals galaxies (squares with dashed black line) broken down into star-forming (circles with dashed blue line) and quenched (circles with dashed red line) galaxies in each of the panels. While the SMF of star-forming galaxies closely trace the central's SMF, quenched galaxies show a drop in the number density at low-masses in agreement with \citetalias{Peng2012}.}\label{fig:SMF_Q_cen_allmodels}

\end{figure*}

\begin{figure*}[t]
\centering
\includegraphics[width=\columnwidth, clip]{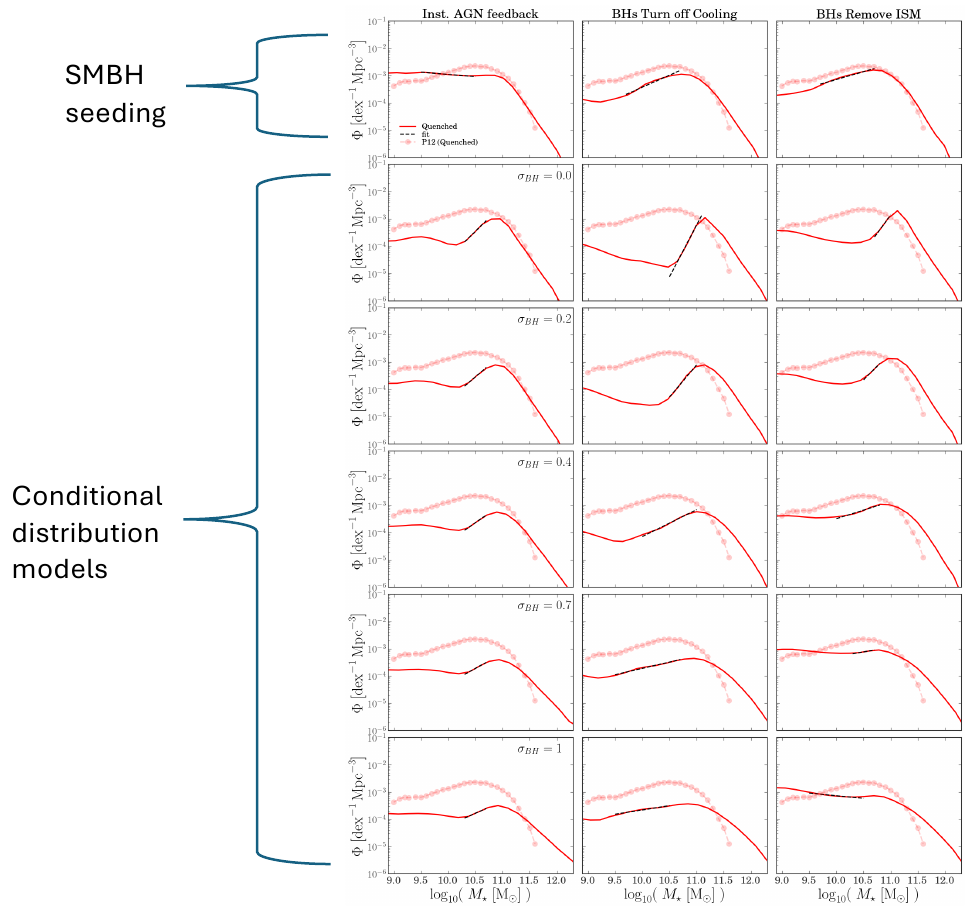}
\caption{Similar to figure \ref{fig:SMF_Q_cen_allmodels}, but here we only show quenched galaxies for {\sc Dark Sage} modified models. The black dashed line shows the low-end fit to obtain the slope $\alpha$. We highlight differences in $\alpha$ with varying BH growth and quenching models.}\label{fig:SMF_Q_cen_allmodels_fit}

\end{figure*}

Figure \ref{fig:SMF_Q_cen_allmodels} shows the SMF of central star-forming and quenched galaxies for five different $\sigma_{\rm BH}$ values, an extended version of Figure \ref{fig:SMF_SFQuenched_models_z0}, where we only show the conditional distribution model at $\sigma_{\rm BH}=0.2$. We find that in most cases, the total SMF at the high-mass end converges with the quenched SMF, except for $\sigma_{\rm BH}>0.4$ where the star-forming galaxies have similar broken powerlaws that traces the total SMF from the low-mass to the high-mass end. On these extreme cases, the drop off at the high-mass end is more shallow due to the relatively large range of stellar masses such black holes live in. Figure \ref{fig:SMF_Q_cen_allmodels_fit} illustrates how the $\alpha$ values from the SMBH seeding and the conditional distribution models are obtained in figure \ref{fig:alpha_sigmaBH}. Our faint-end slopes are well-fitted to represent the drop in the number density of central quenched galaxies below their respective characteristic mass.

%TC:endignore
\end{document}